\documentclass[pra,showpacs,twocolumn]{revtex4}
\usepackage[dvips]{graphicx}
\usepackage[usenames]{color}
\usepackage{amssymb}
\usepackage{amsmath}
\usepackage{subfigure}

\begin{document}

\title{Correlation and Entanglement of Multipartite States}

\author{Y. B. Band and I. Osherov}
\affiliation{Departments of Chemistry and Electro-Optics and the Ilse
Katz Center for Nano-Science, \\
Ben-Gurion University, Beer-Sheva 84105, Israel}

\date{\today}

\begin{abstract}
We derive a classification and a measure of classical- and
quantum-correlation of multipartite qubit, qutrit, and in general,
$n$-level systems, in terms of SU$(n)$ representations of density
matrices.  We compare the measure for the case of bipartite
correlation with concurrence and the entropy of entanglement.  The
characterization of correlation is in terms of the number of nonzero
singular values of the correlation matrix, but that of mixed state
entanglement requires additional invariant parameters in the density
matrix.  For the bipartite qubit case, the condition for mixed state
entanglement is written explicitly in terms of the invariant paramters
in the density matrix.  For identical particle systems we analyze the
effects of exchange symmetry on classical and quantum correlation.
\end{abstract}

\pacs{03.67.-a, 03.67.Mn, 03.65.Ud}

\maketitle

Quantum entanglement is an information resource; it plays an important
role in many protocols for quantum-information processing, including
quantum computation \cite{PShor_97_Grover_96}, quantum cryptography
\cite{Ekert_91}, teleportation \cite{Bennett_93}, superdense coding
\cite{Bennett_92}, and quantum error correction protocols \cite{qec}.
Techniques for characterizing the bipartite entanglement and
correlation of pure and mixed quantum states have enabled many
advances in quantum information and the study of decoherence
\cite{White_01}.  Many quantum information protocols use bipartite
entanglement, but multipartite entanglement \cite{GHZ_89}, also has
quantum-information applications, e.g., controlled secure direct
communication \cite{cont-com}, quantum error correction
\cite{Calderbank_96}, controlled teleportation \cite{Karlsson_98} and
secret sharing \cite{Hillery_99}.  It has been shown that any
inseparable two-qubit states can be distilled to a singlet-state form
with enough copies of the qubit-pairs \cite{Horodecki_97} and
algorithms for multi-copy entanglement distillation for pairs of
qubits have been developed \cite{Dehaene}.  Moreover, multipartite
entanglement offers a means of enhancing interferometric precision
beyond the standard quantum limit and is therefore relevant to
increasing the precision of atomic clocks by decreasing projection
noise in spectroscopy \cite{Wineland_94}.  Here we use a
representation of the density matrix for qubit, qutrit, and more
generally, $n$-level systems containing 2, 3, \ldots, and $N$-parts,
in terms of the correlations between the subsystems to quantify the
classical and quantum correlation of multipartite systems.  Our
classification of correlation is in terms of the correlation matrix
and its singular values, and our classification of entanglement of
mixed states \cite{Werner_89} is associated with the Peres-Horodecki
criterion \cite{Peres_06}, which we express in terms of additional
invariant parameters in the density matrix.  Separate measures of
bipartite, tripartite, etc., correlation are required, since general
mixed states can have bipartite correlation as well as higher
subsystem-number-correlation.

Werner \cite{Werner_89} defined a mixed state of an $N$-partite system
as {\em separable}, i.e., {\em classically-correlated}, if it can be
written as a convex sum,
\begin{equation}   \label{Eq:1}
  \rho = \sum_k p_k \, \rho^A_k \rho^B_k \ldots \rho^N_k ~, \quad p_k
  > 0 ~, \quad \sum_k p_k = 1 ~,
\end{equation}
where $\rho^A_k$ is a valid density matrix of subsystem $A$, etc.
Otherwise, Werner defined it to be {\em entangled}, i.e., {\em
quantum-correlated}.  Unfortunately, this definition of entanglement
for mixed states is not constructive, since, in general, it cannot be
used to decide whether a given density matrix is separable or
entangled.  Moreover, a quantitative measure of entanglement of
multi-partite systems has proven to be difficult to devise.  Note that
studies of the best separable approximation to an arbitrary density
have been carried out and have led to a proposal of a measure for
entanglement \cite{sep_app}.  Furthermore, aspects of the geometry of
separability and entanglement based on Schmidt decomposition have been
studied and led to an analysis of the question of separability for the
two-qubit case \cite{Leinaas_06}.

In what follows, we categorize classically-correlated and
quantum-correlated states and characterize their correlation in terms
the number of nonzero singular values \cite{lin_alg}, $\{d_i\}$, of
the correlation matrix ${\bf C}$, and characterize entanglement of
bitpartite qubit systems using the Peres-Horodecki criterion
\cite{Peres_06} which is reformulated totally in terms of the
parameters used in forming the density matrix.

First, let us consider a bipartite qubit system.  For two uncorrelated
qubits, call them $A$ and $B$, we can write the density matrix as a
product, $\rho_{AB} = \rho_{A} \rho_{B}$, where the individual qubit
density matrices can be written as $\rho_J = \frac{1}{2} \, (1 + {\bf
n}_J \cdot {\boldsymbol \sigma}_J)$, where $J = A,B$, the
${\boldsymbol \sigma}_J$ are Pauli matrices for particle $J$ and the
Bloch vectors are ${\bf n}_J = \langle {\boldsymbol \sigma}_J \rangle
= \mathrm{Tr} \, {\boldsymbol \sigma}_J \rho_J$ \cite{Fano_83}.  For
two correlated qubits,
\begin{equation} \label{Eq:2}
    \rho_{AB} = \frac{1}{4} \, \left[(1 + {\bf n}_A \cdot {\boldsymbol
    \sigma}_A) \, (1 + {\bf n}_B \cdot {\boldsymbol \sigma}_B) +
    {\boldsymbol \sigma}_A \cdot {\bf C}^{AB} \cdot {\boldsymbol
    \sigma}_B \right] ,
\end{equation}
where the tensor ${\bf C}^{AB}$ specifies the qubit correlations,
\begin{equation} \label{Eq:3}
    C^{AB}_{ij} \equiv \langle \sigma_{i,A} \sigma_{j,B} \rangle - \langle
    \sigma_{i,A} \rangle \langle \sigma_{j,B} \rangle = \langle
    \sigma_{i,A} \sigma_{j,B} \rangle - n_{i,A} \, n_{j,B} .
\end{equation}
The density matrix $\rho_{AB}$ is a 4$\times$4 Hermitian matrix with
trace unity, so 15 parameters are required to parameterize it.  The 3
components of ${\bf n}_A$, the 3 components of ${\bf n}_B$, and the 9
components $C_{ij}$ of the 3$\times$3 matrix {\bf C}, where we have no
longer explicitly shown the subsystem superscripts, are sufficient for
this purpose.

Similarly for the bipartite qutrit case.  The 3$\times$3 density
matrix of a single qutrit can be written as $\rho = \frac{1}{3} \,
\left(1 + \frac{3}{2} \langle \lambda_{i} \rangle \lambda_{i}\right)$
where the $\lambda_i$ are the eight traceless Hermitian Gellman
matrices familiar from SU(3) \cite{Georgi_99}, and $\langle
\lambda_{i} \rangle = \mathrm{Tr} \, \lambda_{i} \rho$.  A bipartite
qutrit density matrix can be parameterized in the form
\begin{equation} \label{Eq:4}
    \rho_{AB} = \frac{1}{9} \, [(1 + \frac{3}{2} \langle
    \lambda_{i,A} \rangle \lambda_{i,A} ) \, (1 +
    \frac{3}{2} \langle \lambda_{j,B} \rangle \lambda_{j,B} ) +
    \frac{9}{4} \lambda_{i,A} C_{ij} \lambda_{j,B} ] ,
\end{equation}
\begin{equation} \label{Eq:5}
    C_{ij} \equiv \langle \lambda_{i,A} \lambda_{j,B} \rangle - \langle
    \lambda_{i,A} \rangle \langle \lambda_{j,B} \rangle ~,
\end{equation}
where $C_{ij}$ specifies the correlation between $\lambda_{i,A}$ and
$\lambda_{j,B}$.  Here, $\rho_{AB}$ is a 9$\times$9 Hermitian matrix
with trace unity, so 80 parameters are required to parameterize it.
The eight components of $\langle \lambda_{i,A} \rangle$, eight
components of $\langle \lambda_{i,B} \rangle$, and 64 components
$C_{ij}$ of the 8$\times$8 matrix ${\bf C}$ are sufficient for this
purpose.  The same procedure can be used for bipartite 4-level systems
using the 15 traceless 4$\times$4 Hermitian generator matrices for
SU(4), and bipartite $n$-level systems with the $n^2-1$ traceless
$n$$\times$$n$ Hermitian matrices.  Likewise, a general qubit-qutrit
6$\times$6 density matrix takes the form $\rho_{AB} = \frac{1}{6} \,
[(1 + n_{i,A} \sigma_{i,A} ) \, (1 + \frac{3}{2} \langle \lambda_{j,B}
\rangle \lambda_{j,B} ) + \frac{6}{4} \sigma_{i,A} C_{ij}
\lambda_{j,B}]$ with $C_{ij} \equiv \langle \sigma_{i,A} \lambda_{j,B}
\rangle - \langle \sigma_{i,A} \rangle \langle \lambda_{j,B} \rangle$,
$i = 1, 2, 3$ and $j= 1, \ldots, 8$.

Our bipartite correlation measure for an $n$-level and $m$-level
system is based on the $(n^2-1)$$\times$$(m^2-1)$ correlation matrix
${\bf C}$:
\begin{equation} \label{Eq:E_C}
    {\cal E}_{C} \equiv \frac{n_<^2}{4(n_<^2-1)} \mathrm{Tr} \, {\bf C}
    {\bf C}^T = \frac{n_<^2}{4(n_<^2-1)} \sum_{i,j} C_{ij} C_{ji}^T ~,
\end{equation}
where $n_< = {\mathrm{min}}(n,m)$.  ${\cal E}_{C} = \frac{n_<^2}
{4(n_<^2-1)} \mathrm{Tr} \, (\rho_{AB} - \rho_{A}\rho_{B})^2$ is a
nonnegative real number.  If ${\bf C}$ is a normal matrix
\cite{lin_alg}, $\mathrm{Tr} \, {\bf C} {\bf C}^T$ equals to the sum
of the squares of its eigenvalues, but ${\bf C}$ need not be normal.
${\cal E}_{C}$ is basis-independent; any rotation in Hilbert space
leaves it unchanged.  The normalization factor $n_<^2/[4(n_<^2-1)]$ in
(\ref{Eq:E_C}) is such that the maximum possible value of ${\cal
E}_{C}$ is unity.  ${\cal E}_{C}$ measures both classical- and
quantum-correlation.  This measure of bipartite correlation was
suggested in Ref.~\cite{Schlienz_95} for pure states and $n=m$.

The correlation matrix ${\bf C}$ quantifies the correlation and the
entanglement of bipartite states.  For pure two-qubit states, the
number of nonzero singular values (NSVs) of ${\bf C}$ is zero for
non-entangled states (${\bf C}$ vanishes), and three for entangled
states.  For classically-correlated states with two terms in the sum
[see Eq.~(\ref{Eq:qubit_CC})], only one NSV occurs, two NSVs occur for
three terms, three NSVs occur for four or more terms, and for
entangled (i.e., quantum-correlated) mixed states there are three
NSVs.  These cases are summarized in Fig.~\ref{Fig.two-qubit-class}.
Entangled mixed states can be differentiated from
classically-correlated states with 3 NSVs by applying the
Peres-Horodecki (PH) partial transposition condition \cite{Peres_06}
[which corresponds to changing the sign of $n_{y,B}$ and the matrix
elements $C^{AB}_{iy}$ that multiply $\sigma_{y,B}$ in (\ref{Eq:2}),
and determining whether the resulting $\rho$ is still a genuine
density matrix --- if it is, the state is classically correlated,
i.e., unentangled but correlated] to the density matrices with 3 NSVs.
The {\underline{{\em only}} categories that cannot be distinguished
without use of the PH condition are the mixed-entangled and the
classically correlated states with $\ge 4$ NSVs.

\begin{figure}[!ht]
\includegraphics[width=\columnwidth]{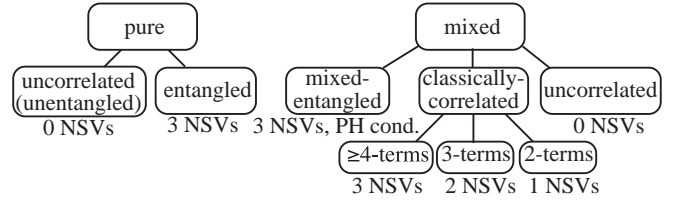}
\caption{Classification of two-qubit states.  Categories can be
experimentally distinguished by measuring ${\bf n}_A$, ${\bf n}_B$,
and using Bell measurements \cite{GHZ_89} to determine the {\bf C}
matrix.}
\label{Fig.two-qubit-class}
\end{figure}

Similarly, for a two qutrit pure state, the number of NSVs of ${\bf
C}$ is zero for non-entangled states (the ${\bf C}$ matrix vanishes),
three, if only two basis states are present in the entangled state,
five, if one of the qutrits contains only two basis states but the
other contains three, and eight if all three basis states are present.
For classically correlated qutrit states, there are 1, 2, \ldots, 8
NSVs for 2, 3, \ldots, and 9 or more terms in the sum, etc.  A similar
classification in terms of the number of NSVs exists for qubit-qutrit
and $n$-level systems.

A general three-qubit density matrix can be written as
\[
    \rho_{ABC} = \frac{1}{8} \, [ (1 + {\bf n}_A \cdot {\boldsymbol
    \sigma}_A) \, (1 + {\bf n}_B \cdot {\boldsymbol \sigma}_B) \, (1 +
    {\bf n}_C \cdot {\boldsymbol \sigma}_C)
\]
\[
    \; \; + {\boldsymbol \sigma}_A \cdot {\bf C}^{AB} \cdot
    {\boldsymbol \sigma}_B + {\boldsymbol \sigma}_A \cdot {\bf C}^{AC}
    \cdot {\boldsymbol \sigma}_C + {\boldsymbol \sigma}_B \cdot {\bf
    C}^{BC} \cdot {\boldsymbol \sigma}_C 
\]
\begin{equation} \label{Eq:6}
    \; \; + \sum_{ijk} \sigma_{i,A}
	\sigma_{j,B} \sigma_{k,C} D_{ijk} ]~,
\end{equation}
where ${\bf C}^{AB}$, ${\bf C}^{AC}$, and ${\bf C}^{BC}$ are the
bipartite correlation matrices and the tensor that specifies the
tripartite correlations is
\begin{equation} \label{Eq:7}
    D_{ijk} \equiv \langle \sigma_{i,A} \sigma_{j,B} \sigma_{k,C} \rangle -
    \langle \sigma_{i,A} \rangle \langle \sigma_{j,B} \rangle \langle
    \sigma_{k,C} \rangle ~.
\end{equation}
A tripartite qutrit state can be similarly parameterized:
\[
    \rho_{ABC} = \frac{1}{27} \, [ \prod_{I} (1 + \frac{3}{2} \sum_i
    \langle \lambda_{i,I} \rangle \lambda_{i,I}) + \frac{9}{4}
    \sum_{I,J} \sum_{i,j} \lambda_{i,I} C_{ij,IJ} \lambda_{j,J}
\]
\begin{equation} \label{Eq:8}
    \; \;  + \frac{27}{8}  \sum_{I,J,K} \sum_{i,j,k} \lambda_{i,A}
	\lambda_{j,B} \lambda_{k,C} D_{ijk,IJK} ]~,
\end{equation}
\begin{equation} \label{Eq:9}
    D_{ijk,IJK} \equiv \langle \lambda_{i,I} \lambda_{j,J} \lambda_{k,K}
    \rangle - \langle \lambda_{i,I} \rangle \langle \lambda_{j,J}
    \rangle \langle \lambda_{k,K} \rangle ~.
\end{equation}

Our tripartite correlation measure ${\cal E}_D$ is based on the
correlation matrix ${\bf D}$, ${\cal E}_{D} \equiv K \sum_{i,j,k}
D_{ijk}^2$, which can also be written as
\begin{equation} \label{Eq:E_D}
    {\cal E}_{D} = K \, \mathrm{Tr} \, (\rho_{ABC} - \rho_{A}\rho_{B}
    \rho_{C} - \sum_{I,J(I \ne J)} \sum_{i,j} C_{ij}^{I,J}
    \sigma_{i,I} \sigma_{j,J} )^2~, 
\end{equation}
where $K = 1/4$ for qubits, and $K = 27/160$ for qutrits with
$\sigma$s replaced by $\lambda$s.  ${\cal E}_D$ is also a
basis-independent nonnegative real number; any rotation in Hilbert
space leaves it unchanged.  A tripartite system may have bipartite- as
well as tripartite-correlation.  The bipartite correlation of a
tripartite system is the sum of the correlation for the three
bipartite pairs,
\begin{equation} \label{Eq:E_C_3}
    {\cal E}_{C} \equiv \frac{n^2}{4(n^2-1)} \sum_{I,J (I \ne J)}
    \mathrm{Tr} \, {\bf C}^{I,J} ({\bf C}^{I,J})^T ~,
\end{equation}
where $I,J = A, B, C$.  The density matrices of four-particle and
higher qubit, qutrits, and $n$-level system states can be constructed
similarly, but with increased complexity.  For example, it is clear
from Eq.~(\ref{Eq:E_D}) how to generalize and obtain the four-particle
correlation of four-particle systems: ${\cal E}_{E} \equiv K'
\sum_{i,j,k,l} E_{ijkl}^2$, where the four-particle-correlation term
of the four-qubit density matrix $\rho_{ABCD}$ is $\sum_{ijkl}
\sigma_{i,A} \sigma_{j,B} \sigma_{k,C} \sigma_{l,D} E_{ijkl}$ and $K'
= 1/8$.

We now present some examples of qubit and qutrit bipartite and
tripartite correlated states.  The maximally entangled bipartite qubit
states are the Bell states,
\begin{equation} \label{Eq:Bell_States}
    |\Psi^{\pm} \rangle \!  = \!  \frac{1}{\sqrt{2}} [ |\!  \!
    \uparrow \downarrow \rangle \pm |\!  \!  \downarrow\uparrow
    \rangle ] ~, \quad |\Phi^{\pm}\rangle \!  = \!  \frac{1}{\sqrt{2}}
    [ |\!  \!  \uparrow \uparrow \rangle \pm |\!  \!  \downarrow
    \downarrow \rangle ] ~.
\end{equation}
For all these states, $\langle
{\boldsymbol \sigma}_{A} \rangle = \langle {\boldsymbol \sigma}_{B}
\rangle = {\bf 0}$, i.e., ${\bf n}_A = {\bf n}_B = {\bf 0}$.  For the
singlet, $\langle \sigma_{i,A} \sigma_{j,B} \rangle = -\delta_{ij}$
(the spins are oppositely polarized).  The density matrices of the
Bell states are:
\begin{eqnarray} \label{Eq:Bell_States_rho}
\rho_{\Psi^{-}} &=& \frac{1}{4} \, (1_A 1_B - {\boldsymbol
\sigma}_A \cdot {\boldsymbol \sigma}_B) ~, \nonumber \\
\rho_{\Psi^{+}} &=& \frac{1}{4} \left( 1_A 1_B + {\boldsymbol
\sigma}_A \cdot {\boldsymbol \sigma}_B - 2\sigma_{z,A} \sigma_{z,B}
\right) ~, \nonumber \\
\rho_{\Phi^{+}} &=& \frac{1}{4} \left( 1_A 1_B + {\boldsymbol
\sigma}_A \cdot {\boldsymbol \sigma}_B - 2\sigma_{y,A} \sigma_{y,B}
\right) ~, \nonumber \\
\rho_{\Phi^{-}} &=& \frac{1}{4} \left( 1_A 1_B + {\boldsymbol
\sigma}_A \cdot {\boldsymbol \sigma}_B - 2\sigma_{x,A} \sigma_{x,B}
\right) ~.
\end{eqnarray}
The correlation matrices of the Bell's states are diagonal and the
correlation measure is ${\cal E}_C=1$, i.e., they are maximally
entangled.

Let us now consider the Rashid pure states \cite{Rashid_78},
\begin{equation} \label{Eq:Rashid_States}
    |\phi^+\rangle = (2 \mathrm{cosh}(2\theta))^{-1/2} \, (e^{-\theta}
    |\!\!\uparrow \uparrow \rangle + e^{\theta} |\!\!\downarrow
    \downarrow \rangle) ~,
\end{equation}
whose density matrix is $\rho_{\phi^+} = \frac{1}{4} \, \{ [1_A -
\tanh(2\theta) \sigma_{z,A}] [1_B - \tanh(2\theta) \sigma_{z,B}] +
\mathrm{sech}(2\theta) ( \sigma_{x,A} \sigma_{x,B} - \sigma_{y,A}
\sigma_{y,B} ) + \mathrm{sech}^2(2\theta) \sigma_{z,A} \sigma_{z,B}
\}$.  When $\theta = 0$, $|\phi^+\rangle = |\Phi^+\rangle$, and as
$\theta \to \pm \infty$, an unentangled state results.  The
nonvanishing correlation matrix elements are: $C_{xx} =
\mathrm{sech}(2\theta), C_{yy} = -\mathrm{sech}(2\theta), C_{zz} =
\mathrm{sech}^2(2\theta)$.  Using (\ref{Eq:E_C}) we obtain the
correlation measure ${\cal E}_C(|\phi^+ \rangle) = \frac{1}{3}
\mathrm{Tr}\, {\bf C} {\bf C}^T = \frac{1}{3}(2 \,
\mathrm{sech}^2(2\theta) + \mathrm{sech}^4(2\theta))$.  The
concurrence ${\cal C}$ \cite{Wootters_98,Peres_06} is $$ {\cal
C}(|\phi^+ \rangle) = \sqrt{2\left(1-\mathrm{Tr} \,
\left[\rho _A{}^2\right]\right)} = \mathrm{sech}(2\theta) ~,$$
since 
$$\rho_A = \frac{1}{2} \, \left(\!  \!  \begin{array}{cc}
1-\mathrm{tanh}(2\theta) & 0 \\ 0 &
1+\mathrm{tanh}(2\theta) \end{array} \!  \!  \right) ~,$$ 
and the entanglement entropy is $S \equiv - \mathrm{Tr} \, [\rho_A
\log_2 \rho_A]$.  These results are graphically presented in
Fig.~\ref{Fig.Corr_measures}.  All the measures equal unity for
$\theta = 0$ and decrease rapidly vs.~$\theta$.

\begin{figure}[!ht]
\includegraphics[width=\columnwidth]{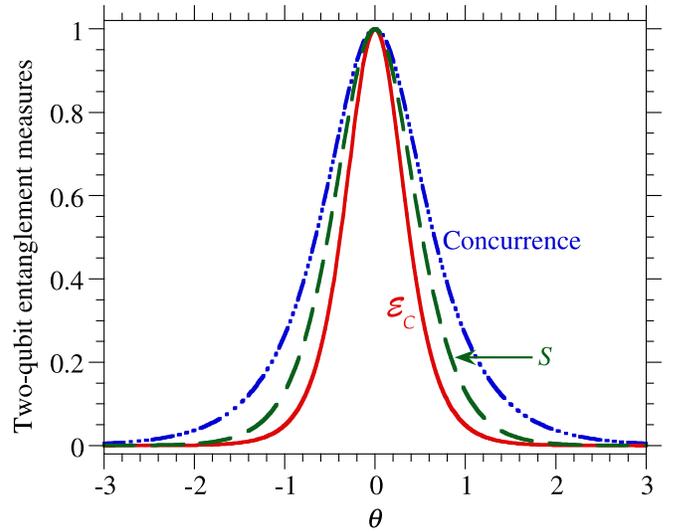}
\caption{(color online) Comparison of the ${\cal E}_C$ measure of
correlation, the concurrence ${\cal C}$ and the entanglement entropy 
$S$ for the Rashid pure states.}
\label{Fig.Corr_measures}
\end{figure}

Two-qubit classically-correlated states take the form
\begin{equation} \label{Eq:qubit_CC}
    {\rho}^{\mathrm{CC}} = \frac{1}{4} \sum_{k \ge 2} p_k \, (1 +
    {\bf n}_{A,k} \cdot {\boldsymbol \sigma}_A) \, (1 + {\bf n}_{B,k}
    \cdot {\boldsymbol \sigma}_B) ~,
\end{equation}
with $\sum_k p_k = 1$ and $p_k > 0$.  The density matrix for the
classically-correlated state can be written in the form of
Eq.~(\ref{Eq:2}) with Bloch vectors
\begin{equation} \label{Eq:qubit_CC_1}
  {\bf n}_{A} = \sum_k p_k \, {\bf n}_{A,k} ~, \quad {\bf n}_{B} =
  \sum_k p_k \, {\bf n}_{B,k} ~,
\end{equation}
and correlation matrix
\begin{equation}  \label{Eq:qubit_CC_2}
    C_{ij} = \sum_k p_k \, n_{i,A,k} \left[ n_{j,B,k} - \sum_l p_l
\, n_{j,B,l} \right] ~.
\end{equation}
For example, for classically-correlated mixed states of the form
$\rho^{\mathrm{CC}} = (2 \, \mathrm{sech}^2(2\theta))^{-1/2}
(e^{-\theta} |\!  \!  \downarrow \uparrow \rangle \, \langle
\downarrow \uparrow \!  \!  |+ e^{\theta}|\!\!\uparrow
\downarrow\rangle \, \langle\uparrow \downarrow \!  \!  |)$, we find
that all the correlation coefficients vanish, except for $C_{zz} = -
\mathrm{sech}^2(2\theta)$, the density matrix in representation
(\ref{Eq:2}) is $\rho^{\mathrm{CC}} = \frac{1}{4} \left( 1_A 1_B -
\mathrm{sech}^2(2\theta) \, \sigma_{z,A} \sigma_{z,B}\right)$, and the
classical-correlation measure is ${\cal E}_C^{\mathrm{CC}} =
\frac{1}{3} \,\mathrm{sech}^4(2\theta)$.

It is elucidating to consider the Werner two-qubit density matrix
composed of a sum of a singlet state and the maximally mixed state,
$\rho^{W} = p |\Psi^{-} \rangle \langle \Psi^{-} | + \frac{1-p}{4}
{\bf 1}$, or, the more general Werner two-qubit density matrix,
\begin{equation}  \label{Eq:GW}
    \rho^{GW} = p \, |\psi^- \rangle \langle \psi^- | + \frac{1-p}{4}
    {\bf 1} ~,
\end{equation}
where $|\psi^- \rangle = (2 \mathrm{cosh}(2\theta))^{-1/2} \,
(e^{-\theta} |\!\!\uparrow \downarrow \rangle - e^{\theta}
|\!\!\downarrow \uparrow \rangle)$.  $\rho^{GW}$ reduces to $\rho^{W}$
for $\theta = 0$.  For $\rho^{GW}$,
\begin{equation}  \label{Eq:GW_n}
    {\bf n}_A = - {\bf n}_B = p \, \mathrm{tanh}(2\theta) \, {\hat{\bf
    z}} ~,
\end{equation}
and
\begin{equation} \label{Eq:14}
    {\bf C}^{GW} = -p \left( \! \!
    \begin{array}{ccc}
      {\mathrm{sech}(2\theta)} & 0 & 0 \\
      0 & {\mathrm{sech}(2\theta)} & 0 \\
       0 & 0 & {1-p + p \, \mathrm{sech}^2(2\theta)}
    \end{array} \! \! \right). 
\end{equation}
The PH entanglement criterion \cite{Peres_06} shows that this state is
entangled if $p [(1 + 2 \mathrm{sech}(2 \theta)] \ge 1$.  Figure
\ref{Ec_vs_p_theta} plots the PH criterion limit and the correlation
measure, ${\cal E}_C(p,\theta) = \sum_i d_i^2 = 1 - p + (2p^2+p)
\mathrm{sech}^2(2\theta)$, for the generalized Werner state.  Note that
the PH criterion is not obtainable from ${\bf C}$ alone, but can be
obtained using the invariant parameters $\xi \equiv \sum_i d_i -
\frac{{\bf n}_A \cdot {\bf C} \cdot {\bf n}_B}{{\bf n}_A \cdot {\bf
n}_B}$ and ${\bf n}_A \cdot {\bf n}_B$.  More explicitly, $p [1 + 2
\mathrm{sech}(2 \theta)] = -\xi + \sqrt{\xi^2/4 - {\bf n}_A \cdot {\bf
n}_B}$, so the PH condition reads
\begin{equation}  \label{Eq:PH}
    -\frac{\xi}{2} + \frac{-\xi + \sqrt{\xi^2 - 4 \, {\bf n}_A \cdot
    {\bf n}_B}}{2} \ge 1 ~,
\end{equation}
which can be written as the condition: the largest root of the
quadratic equation, $(x + \xi/2)^2 + \xi (x + \xi/2) + {\bf n}_A \cdot
{\bf n}_B = 0$, is greater than unity.  Thus, mixed state entanglement
is determined not only by ${\bf C}$ but by additional invariant
characteristics of the density matrix, i.e., invariant characteristics
composed of the parameters ${\bf C}$, ${\bf n}_A$ and ${\bf n}_B$ used
to form the density matrix (whereas the correlation is determined only
in terms of ${\bf C}$).  The physical significance of the scalar
product ${\bf n}_A \cdot {\bf n}_B$ as the projection of the
expectation value of the spin of one qubit on the other, is clear, as
is the physical significance of $\xi$ as a specific projection of the
singular values of the correlation matrix that depends on the average
spins ${\bf n}_A$ and ${\bf n}_B$ \cite{xi}.  However, the physical
significance of the PH entanglement criterion is not yet clear; i.e.,
the physical interpretation of Eq.~(\ref{Eq:PH}) [or the quadratic
equation] remains to be uncovered.  But at least the PH condition is
now expressed only in terms of the physical parameters appearing in
the density matrix, rather than by the partial transposition
condition, which is more removed from physical interpretation.

\begin{figure}[!ht]
\includegraphics[width=\columnwidth]{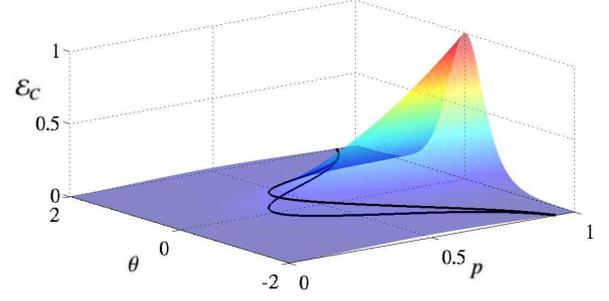}
\caption{(color online) ${\cal E}_C(p,\theta)$ versus $p$ and $\theta$
for the generalized Werner density matrix $\rho^{GW}$, and the
Peres-Horodecki entanglement criterion limit, $p [1 + 2
\mathrm{sech}(2 \theta)] = 1$, drawn on the $p$-$\theta$ plane and
projected onto the ${\cal E}_C$ surface.}
\label{Ec_vs_p_theta}
\end{figure}

As an example of a tripartite pure qutrit state, consider
\begin{equation} \label{Eq:3_qutrit_1}
     |\psi^{E3}\rangle = \frac{e^{\theta_1} e^{\theta_2} | v_1 v_1 v_1
     \rangle + e^{-\theta_1} | v_2 v_2 v_2 \rangle + e^{-\theta_2} |
     v_3 v_3 v_3 \rangle}{\sqrt{e^{2\theta_1} e^{2\theta_2} +
     e^{-2\theta_1} + e^{-2\theta_2}}} ,
\end{equation}
where $| v_1 \rangle = (1,0,0)^T$, $| v_2 \rangle = (0,1,0)^T$, $| v_3
\rangle = (0,0,1)^T$.  The qutrit bipartite correlation ${\cal E}_C$
and tripartite correlation ${\cal E}_D$ are plotted in
Fig.~\ref{Fig3.3_qutrit_ED_EC}.  The maximum of ${\cal E}_D$ is at
$\theta_1 = \theta_2 = 0$, where ${\cal E}_D = 1$, but the bipartite
correlation dips there.  Three ridges of high bipartite and tripartite
correlation occur, one at $\theta_1 = \theta_2 =$ negative, and two
others at 120 degrees rotation from the first.

\begin{figure}[!ht]
\centering\subfigure[]{\includegraphics[width=\columnwidth]{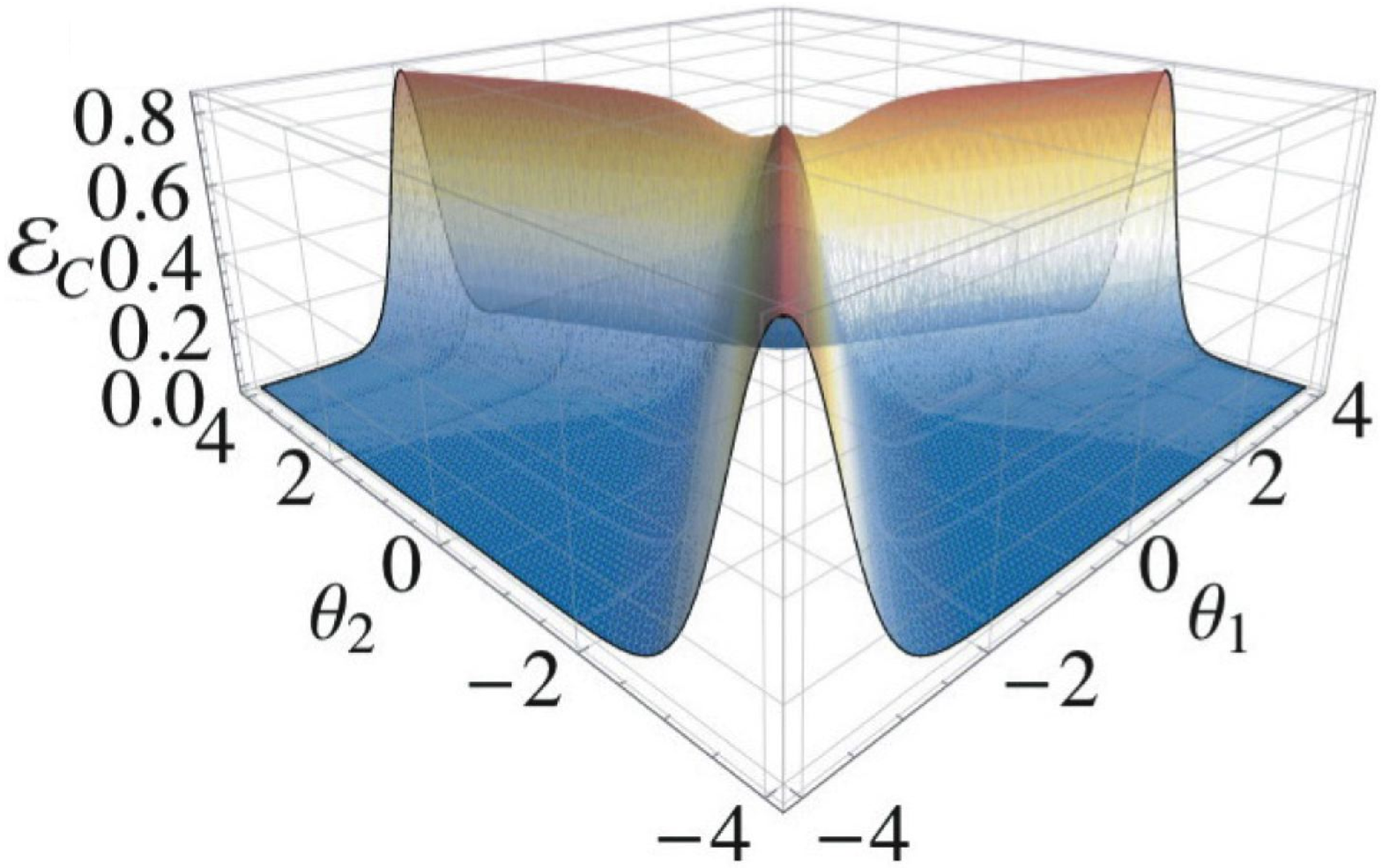}}
\centering\subfigure[]{\includegraphics[width=\columnwidth]{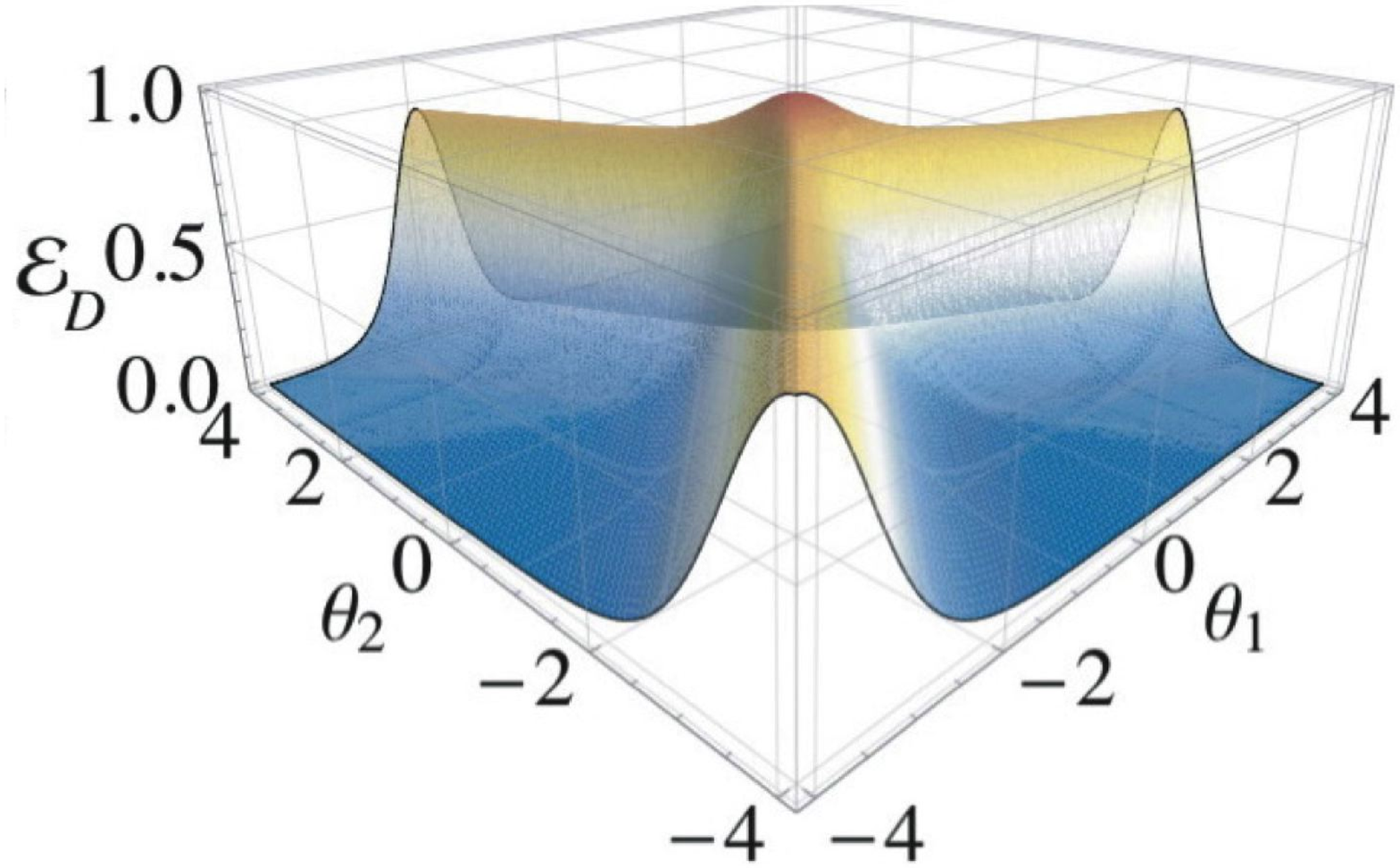}}
\caption{(color online) The bipartite and tripartite correlation
measures, ${\cal E}_C$ and ${\cal E}_D$ for the tripartite-qutrit pure
state (\ref{Eq:3_qutrit_1}).}
\label{Fig3.3_qutrit_ED_EC}
\end{figure}

In most quantum information systems, qubits are distinguishable, so
there is no need to account for exchange symmetry, but for {\em
identical} bosonic or fermionic systems, the density matrix must be
properly symmetrized, e.g., $\rho_{AB}^{\mathrm{sym}} = {\cal S}
\rho_{AB} {\cal S}$.  For two qubits, the symmetrization operator is
${\cal S} = \frac{1}{2} (1 + P_{AB}) = \frac{3}{4} +
\frac{1}{4}{\boldsymbol \sigma}_A \cdot {\boldsymbol \sigma}_B$ and
the antisymmetrization operator is ${\cal A} = \frac{1}{2} (1 -
P_{AB}) = \frac{1}{4} - \frac{1}{4}{\boldsymbol \sigma}_A \cdot
{\boldsymbol \sigma}_B$, hence,
$\rho_{AB}^{\mathrm{sym}} = \left(\! \frac{3}{4} +
    \frac{1}{4}{\boldsymbol \sigma}_A \cdot {\boldsymbol \sigma}_B
    \! \right) \rho_{AB} \left(\! \frac{3}{4} + \frac{1}{4}{\boldsymbol
    \sigma}_A \cdot {\boldsymbol \sigma}_B \! \right)$,
$\rho_{AB}^{\mathrm{anti}} = \left(\! \frac{1}{4} -
    \frac{1}{4}{\boldsymbol \sigma}_A \cdot {\boldsymbol \sigma}_B
    \! \right) \rho_{AB} \left(\! \frac{1}{4} - \frac{1}{4}{\boldsymbol
    \sigma}_A \cdot {\boldsymbol \sigma}_B \! \right)$.
If the qubits are antisymmetric, $\rho_{AB} =
\rho_{AB}^{\mathrm{anti}} = {\cal A} \rho_{AB} {\cal A}$, the state
must be pure singlet, $\rho_{AB}^{\mathrm{anti}} = \frac{1}{4} (1 -
{\boldsymbol \sigma}_{A} \cdot {\boldsymbol \sigma}_{B})$; it cannot
be a mixed state, as opposed to $\rho_{AB}^{\mathrm{sym}}$ which can
be mixed.  If spatial degrees of freedom need to be included in the
description, in addition to the internal degrees of freedom, a
bipartite density matrix can always be written as a product of an
internal (i.e., spin) part and an external (i.e., space) part.  Hence,
a symmetric density matrix $\rho_{AB}^{\mathrm{sym}}$ for the internal
degrees of freedom must be multiplied by a symmetric [antisymmetric]
spatial density matrix for the spatial degrees of freedom $\{{\bf
r}_A$, ${\bf r}_B\}$ for bosons [fermions], and
$\rho_{AB}^{\mathrm{anti}}$ must be multiplied by an antisymmetric
[symmetric] spatial density matrix for bosons [fermions], so that the
full density matrix has the right exchange symmetry.  We show
elsewhere that this has relevance to collisional shifts in atomic
clocks \cite{YB_IO_10}.

In summary, we have developed a classification of correlation for
multipartite $N$-level quantum systems by writing their density
matrices in terms of SU$(N)$ generators, and we defined a measure of
correlation for such systems, based upon their correlation matrices.
The entanglement involves not just the correlation matrix but also
other invariant parameters in the density matrix for the system.  This
formulation can now be used for a variety of applications, e.g., in
the optimization of quantum gates and in the calculation of
collisional clock shifts.

This work was supported in part by grants from the U.S.-Israel
Binational Science Foundation (No.~2006212), the Israel Science
Foundation (No.~29/07), and the James Franck German-Israel Binational
Program.

\end{document}